\documentclass[twoside,twocolumn,9pt]{article}
\usepackage{extsizes}
\usepackage[super,sort&compress,comma]{natbib} 
\usepackage[version=3]{mhchem}
\usepackage[left=1.5cm, right=1.5cm, top=1.785cm, bottom=2.0cm]{geometry}
\usepackage{balance}
\usepackage{times,mathptmx}
\usepackage{amssymb}
\usepackage{amsmath}
\usepackage{braket}
\usepackage{sectsty}
\usepackage{graphicx} 
\usepackage{lastpage}
\usepackage[format=plain,justification=justified,singlelinecheck=false,font={stretch=1.125,small,sf},labelfont=bf,labelsep=space]{caption}
\usepackage{float}
\usepackage{fancyhdr}
\usepackage{fnpos}
\usepackage{color,soul}
\usepackage[english]{babel}
\addto{\captionsenglish}{%
  
}
\usepackage{array}
\usepackage{droidsans}
\usepackage{charter}
\usepackage[T1]{fontenc}
\usepackage[usenames,dvipsnames]{xcolor}
\usepackage{setspace}
\usepackage[compact]{titlesec}
\usepackage{hyperref}
\usepackage{xcolor}
\usepackage{soul}
\usepackage[section]{placeins}
\usepackage{dblfloatfix}    % To enable figures at the bottom of page

\usepackage{epstopdf}%This line makes .eps figures into .pdf - please comment out if not required.

\definecolor{cream}{RGB}{222,217,201}

\begin{document}

\pagestyle{fancy}
\thispagestyle{plain}
\fancypagestyle{plain}{

%%%HEADER%%%
\fancyhead[C]{\includegraphics[width=18.5cm]{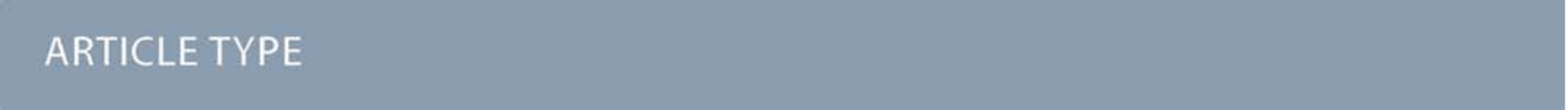}}
\fancyhead[L]{\hspace{0cm}\vspace{1.5cm}\includegraphics[height=30pt]{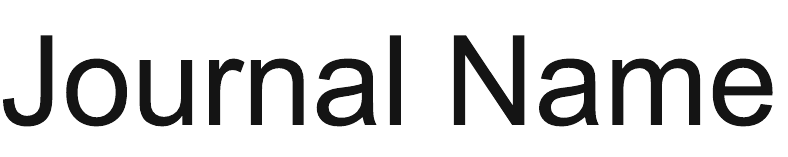}}
\fancyhead[R]{\hspace{0cm}\vspace{1.7cm}\includegraphics[height=55pt]{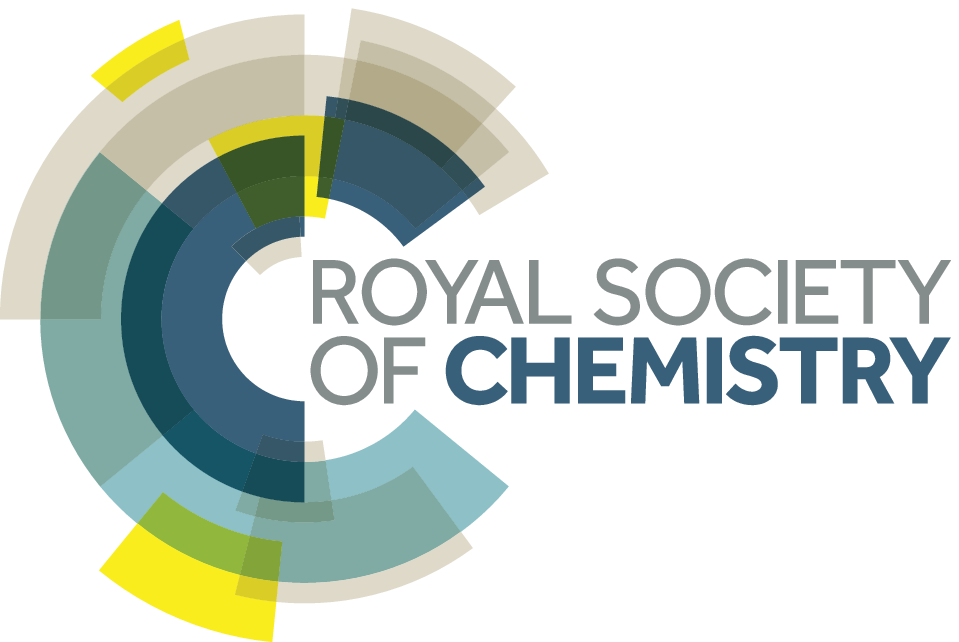}}
\renewcommand{\headrulewidth}{0pt}
}
%%%END OF HEADER%%%

%%%PAGE SETUP - Please do not change any commands within this section%%%
\makeFNbottom
\makeatletter
\renewcommand\LARGE{\@setfontsize\LARGE{15pt}{17}}
\renewcommand\Large{\@setfontsize\Large{12pt}{14}}
\renewcommand\large{\@setfontsize\large{10pt}{12}}
\renewcommand\footnotesize{\@setfontsize\footnotesize{7pt}{10}}
\makeatother

\renewcommand{\thefootnote}{\fnsymbol{footnote}}
\renewcommand\footnoterule{\vspace*{1pt}% 
\color{cream}\hrule width 3.5in height 0.4pt \color{black}\vspace*{5pt}} 
\setcounter{secnumdepth}{5}

\makeatletter 
\renewcommand\@biblabel[1]{#1}            
\renewcommand\@makefntext[1]% 
{\noindent\makebox[0pt][r]{\@thefnmark\,}#1}
\makeatother 
\renewcommand{\figurename}{\small{Fig.}~}
\sectionfont{\sffamily\Large}
\subsectionfont{\normalsize}
\subsubsectionfont{\bf}
\setstretch{1.125} %In particular, please do not alter this line.
\setlength{\skip\footins}{0.8cm}
\setlength{\footnotesep}{0.25cm}
\setlength{\jot}{10pt}
\titlespacing*{\section}{0pt}{4pt}{4pt}
\titlespacing*{\subsection}{0pt}{15pt}{1pt}
%%%END OF PAGE SETUP%%%

%%%FOOTER%%%
\fancyfoot{}
\fancyfoot[LO,RE]{\vspace{-7.1pt}\includegraphics[height=9pt]{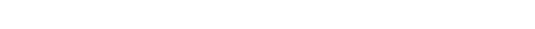}}
\fancyfoot[CO]{\vspace{-7.1pt}\hspace{13.2cm}\includegraphics{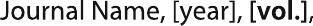}}
\fancyfoot[CE]{\vspace{-7.2pt}\hspace{-14.2cm}\includegraphics{head_foot/RF}}
\fancyfoot[RO]{\footnotesize{\sffamily{1--\pageref{LastPage} ~\textbar  \hspace{2pt}\thepage}}}
\fancyfoot[LE]{\footnotesize{\sffamily{\thepage~\textbar\hspace{3.45cm} 1--\pageref{LastPage}}}}
\fancyhead{}
\renewcommand{\headrulewidth}{0pt} 
\renewcommand{\footrulewidth}{0pt}
\setlength{\arrayrulewidth}{1pt}
\setlength{\columnsep}{6.5mm}
\setlength\bibsep{1pt}
%%%END OF FOOTER%%%

%%%FIGURE SETUP - please do not change any commands within this section%%%
\makeatletter 
\newlength{\figrulesep} 
\setlength{\figrulesep}{0.5\textfloatsep} 

\newcommand{\topfigrule}{\vspace*{-1pt}% 
\noindent{\color{cream}\rule[-\figrulesep]{\columnwidth}{1.5pt}} }

\newcommand{\botfigrule}{\vspace*{-2pt}% 
\noindent{\color{cream}\rule[\figrulesep]{\columnwidth}{1.5pt}} }

\newcommand{\dblfigrule}{\vspace*{-1pt}% 
\noindent{\color{cream}\rule[-\figrulesep]{\textwidth}{1.5pt}} }

\makeatother
%%%END OF FIGURE SETUP%%%

%%%TITLE, AUTHORS AND ABSTRACT%%%
\twocolumn[
  \begin{@twocolumnfalse}
\vspace{3cm}
\sffamily
\begin{tabular}{m{4.5cm} p{13.5cm} }

\includegraphics{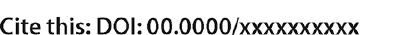} & \noindent\LARGE{\textbf{Full color generation with Fano-type resonant HfO$_\textrm{2}$ nanopillars designed by a deep-learning approach}} \\%Article title goes here instead of the text "This is the title"
\vspace{0.3cm} & \vspace{0.3cm} \\

 & \noindent\large{Omid Hemmatyar,\texttt{$^{\ddag}$} Sajjad Abdollahramezani,\texttt{$^{\ddag}$} Yashar Kiarashinejad, Mohammadreza Zandehshahvar, and Ali Adibi$^{\ast}$} \\%Author names go here instead of "Full name", etc.

\includegraphics{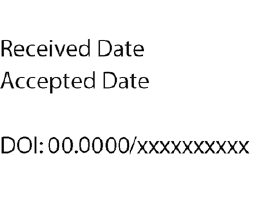} & \noindent\normalsize{In contrast to lossy plasmonic metasurfaces (MSs), wideband dielectric MSs comprising of subwavelength nanostructures supporting Mie resonances are of great interest in the visible wavelength range. Here, for the first time to our knowledge, we experimentally demonstrate a reflective MS consisting of a square-lattice array of Hafnia (HfO$_2$) nanopillars to generate a wide color gamut. To design and optimize these MSs, we use a deep-learning algorithm based on a dimensionality reduction technique. Good agreement is observed between simulation and experimental results in yielding vivid and high-quality colors. We envision that these structures not only empower the high-resolution digital displays and sensitive colorimetric biosensors but also can be applied to on-demand applications of beaming in a wide wavelength range down to deep ultraviolet.} \\%The abstrast goes here instead of the text "The abstract should be..."

\end{tabular}

 \end{@twocolumnfalse} \vspace{0.6cm}

  ]
%%%END OF TITLE, AUTHORS AND ABSTRACT%%%

%%%FONT SETUP - please do not change any commands within this section
\renewcommand*\rmdefault{bch}\normalfont\upshape
\rmfamily
\section*{}
\vspace{-1cm}

%%%FOOTNOTES%%%

\footnotetext{\textit{School of Electrical and Computer Engineering, Georgia Institute of Technology, 778 Atlantic Drive NW, Atlanta, GA 30332, USA; E-mail: ali.adibi@ece.gatech.edu}}

\footnotetext{\ddag~These authors contributed equally to this work.}

%%%END OF FOOTNOTES%%%

%%%MAIN TEXT%%%%

\section*{Introduction}

{C}{olors} play an important role for us to obtain visual information from our surrounding. Instead of generating colors in a conventional manner through absorption or emission of light from chemical pigments of materials, one can get inspired by nature to artificially produce the structural colors \cite{vukusic1999quantified}. In structural coloration, the desired colors are generated through the interaction of light with subwavelength structured materials, which act as appropriate color filters for incident white light. Structural colors are resistant to photobleaching, high temperature, and ultraviolet irradiation; they are not hazardous to the environment, either. As a result, they have attracted significant attention as a promising alternative for conventional pigment- or dye-based colors \cite{kristensen2017plasmonic}.

One way to generate structural colors is to use metasurfaces (MSs), i.e., two dimensional (2D) arrays of patterned nanoresonators with subwavelength features. In recent years, plasmonic and dielectric MSs have been extensively used to create interesting functionalities by locally manipulating the amplitude, phase, polarization, and frequency of the incident light \cite{yu2014flat,zhan2016low,colburn2018metasurface,kwon2018computational,mahmood2018polarisation,taghinejad2019all,taghinejad2018ultrafast,taghinejad2018hot,abdollahramezani2015analog,abdollahramezani2018reconfigurable,hemmatyar2017phase}. The capability of the surface plasmon resonances (i.e., the collective excitation of electrons and electromagnetic waves at the metallic surfaces) to confine the incident light below the diffraction limit enables plasmonic MSs to generate high-resolution structural colors \cite{king2015fano,ding2017gradient,kristensen2017plasmonic,wang2017full,rezaei2017correlation,rezaei2019wide,daqiqeh2019tunable}. Vivid color generation with rich saturation in a large color gamut necessitates MSs with high-quality-factor (or high-$Q$) resonances (i.e., narrow-band resonances) with near-unity intensity in their reflectance or transmittance spectra. However, significant intrinsic losses of metals inevitably used in the plasmonic MSs in the visible range broaden the resonance spectra, and consequently, hinder the realization of high-efficiency structural colors with large color gamut and purity. Furthermore, gold, silver, and aluminum as the most common metals used in plasmonics-based coloration suffer from high cost and/or CMOS-non-compatibility, which hampers the foundry-level production of the resulting MSs \cite{kristensen2017plasmonic,shao2018advanced}. 

To circumvent such challenges, all-dielectric MSs, which support both electric-dipole (ED) and magnetic-dipole (MD) resonances (in contrast to most plasmonic MSs that only provide ED resonances), associated with Mie scattering have been proposed and implemented \cite{staude2013tailoring,zhu2017resonant}. Although both plasmonic and all-dielectric MSs can control the resonance properties by tuning the geometrical parameters \cite{abdollahramezani2015beam,abdollahramezani2017dielectric,chizari2016analog}, the latter provide high-$Q$ resonances as well as lower intrinsic Ohmic losses. Silicon (Si) as a high-refractive-index material in the visible range has been widely used in different patterned geometries such as square \cite{Si_Square}, circular \cite{a_Si_disk,gawlik2018structural,nagasaki2018control,berzins2018color} and cross-shaped \cite{cross} nanopillars (NPs) to generate structural colors. However, Si suffers from high material losses for wavelengths smaller than 600 nm preventing generation of high-quality colors. As a promising alternative, transparent materials with median index of refraction such as amorphous titanium dioxide (TiO$_2$), and low index of refraction such as silicon oxide (SiO$_2$), and polymers (with wide transparency windows in the entire visible range), have been more recently introduced \cite{kim2018outfitting}. Among different options, TiO$_2$ has been used most widely for generating high-quality colors due to its higher refractive index leading to more confined modes \cite{sun2017all,koirala2018highly}. However, there is a growing interest in ultrawide-bandgap median-contrast materials (with refractive indices comparable to TiO$_2$) to extend the use of MSs to the ultraviolet (UV) wavelength range \cite{sss}. Meanwhile, compatibility with well-developed conformal deposition techniques is indispensable to create high-aspect ratio patterned nanoresonators supporting Mie resonances. 

In this paper, for the first time to our knowledge, an all-dielectric MS consisting of a square-lattice of Hafnia (HfO$_2$) NPs for generation of a large high-quality color gamut based on Fano-type resonance is experimentally demonstrated. Due to the high aspect-ratio of constituent dielectric NPs, both ED and MD Mie-type resonances are observed in the reflectance response. On the other hand, the radiation of these Mie-type resonances can strongly be coupled to the direct reflected light. This in turn, leads to the narrow-band Fano-type resonances, which are desirable for generating vivid and high-quality structural colors \cite{shen2015structural}. Moreover, we leverage a novel deep-learning approach based on dimensionality reduction not only in our design methodology, but also in understanding the fundamental mechanism of light-matter interaction in the nanoscale regime.

% \FloatBarrier

\section*{Methods and Results}

\begin{figure}[t]
\centering
\hspace{0.1cm}
\includegraphics[width=1\linewidth, trim={0cm 0cm 0cm 0cm},clip]{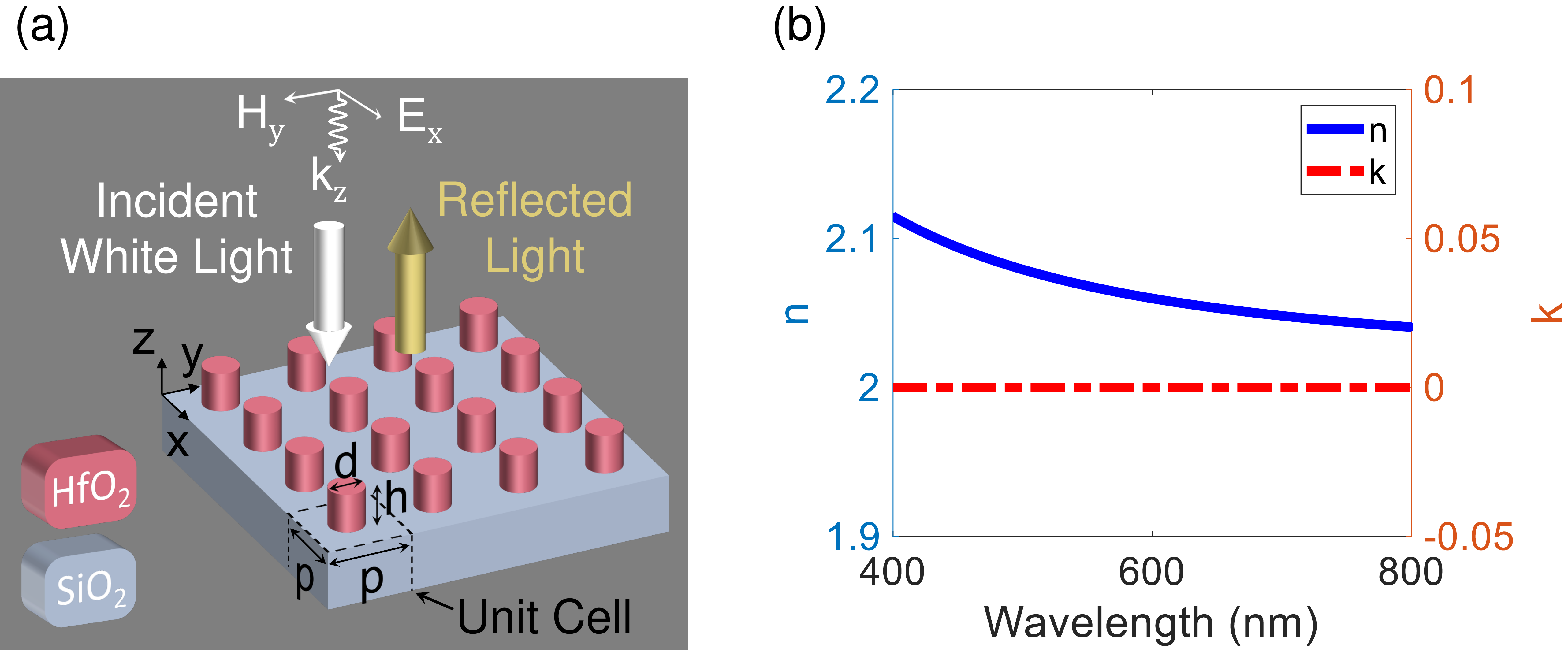}
\caption{(a) Schematic of the all-dielectric MS consisting of HfO$_2$ NPs fabricated on a SiO$_\textrm{2}$ substrate. While the height ($h$) of the NPs is fixed at 350 nm, the diameter is chosen as $d$~=~0.75$p$, where change in periodicity ($p$) results in the generation of different colors. (b) The real part (solid blue line) and imaginary part (dashed red line) of the refractive index of HfO$_2$ deposited by ALD.}
\label{fig:Structure}
\end{figure}

\begin{figure*}[t]
\centering
\hspace{0.1cm}
\includegraphics[width=1\linewidth, trim={0cm 2cm 0cm 2cm},clip]{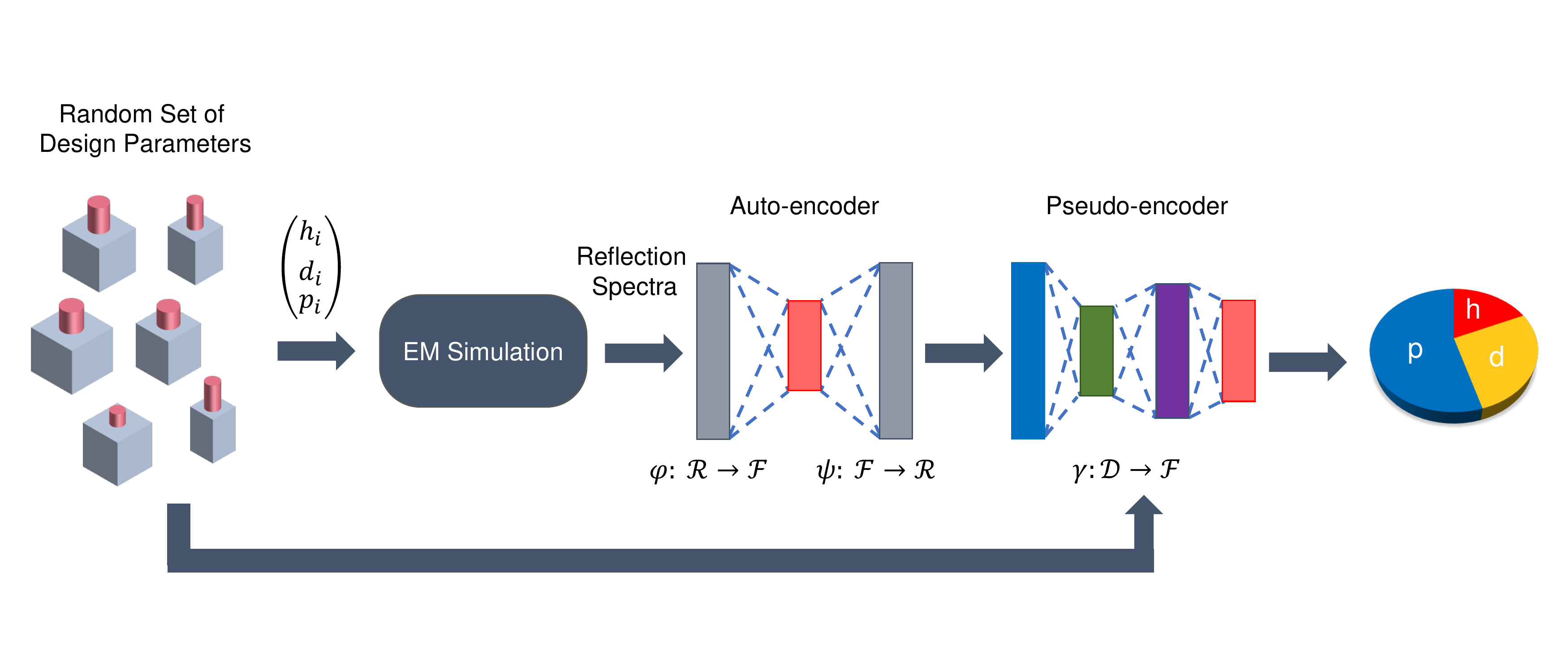}
\caption{The process of determining the effect of variation of each design parameter in changing the reflectance response, and consequently, generated colors. First, a set of MSs with random design parameters (i.e., $h_{i}$, $d_{i}$, and $p_{i}$) are simulated through an EM solver software, and then the reflectance responses are fed into the deep-learning-based approach as the training set. The dimensionality of the reflectance data is reduced through an auto-encoder. Then, the output of the auto-encoder is fed into a pseudo-encoder to extract the importance of each design parameter.}
\label{fig:Blockdiagram}
\end{figure*}

\begin{figure}[!b]
\centering
\hspace{0.1cm}
\includegraphics[width=0.6\linewidth, trim={0cm 0cm 0cm 0cm},clip]{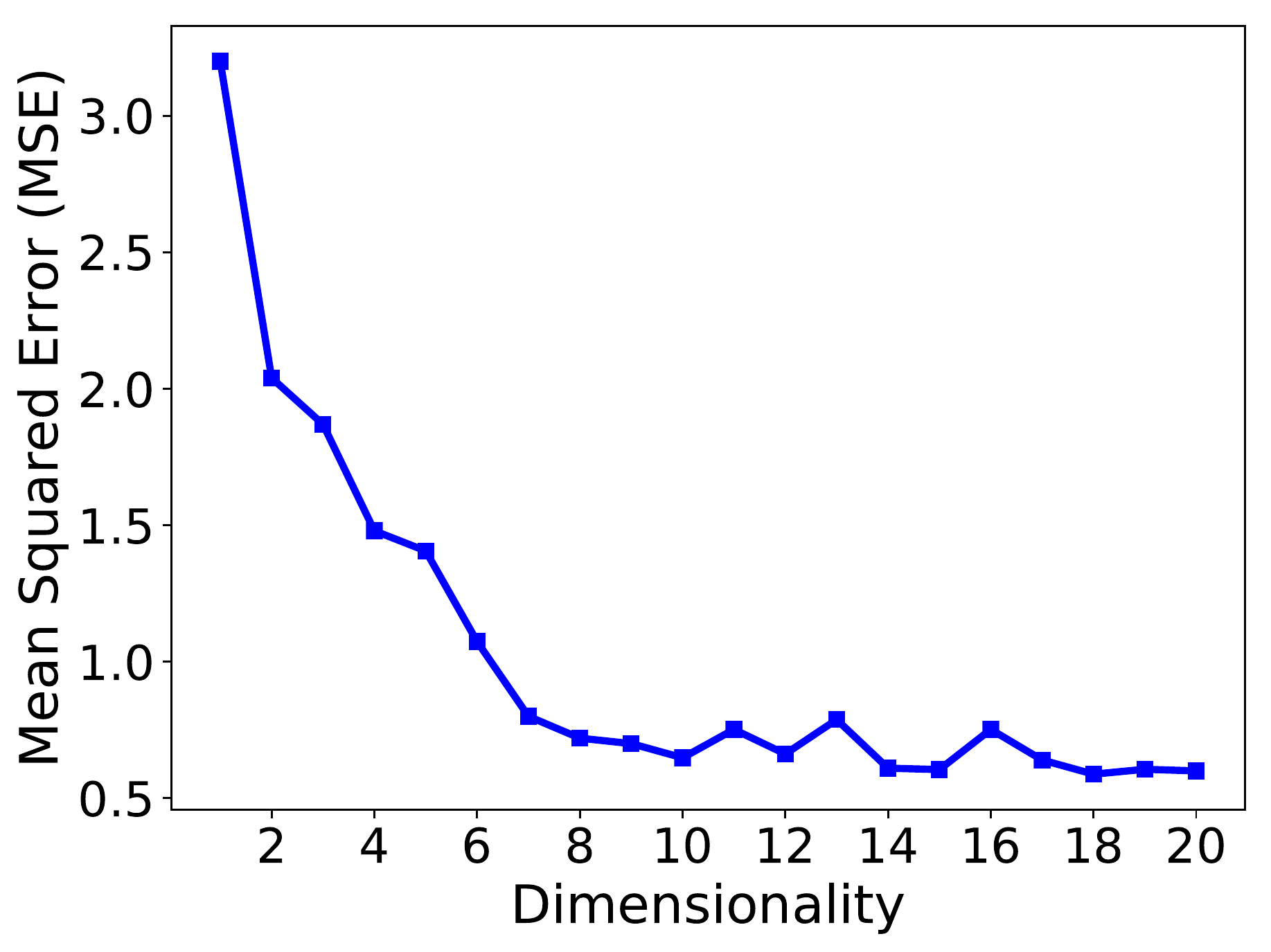}
\caption{MSE of the trained auto-encoder for different dimensionality of the latent space.}
\label{fig:MSE}
\end{figure}
\begin{figure}[!b]
\centering
% \hspace{0.1cm}
\includegraphics[width=1\linewidth, trim={0cm 1cm 0cm 0cm},clip]{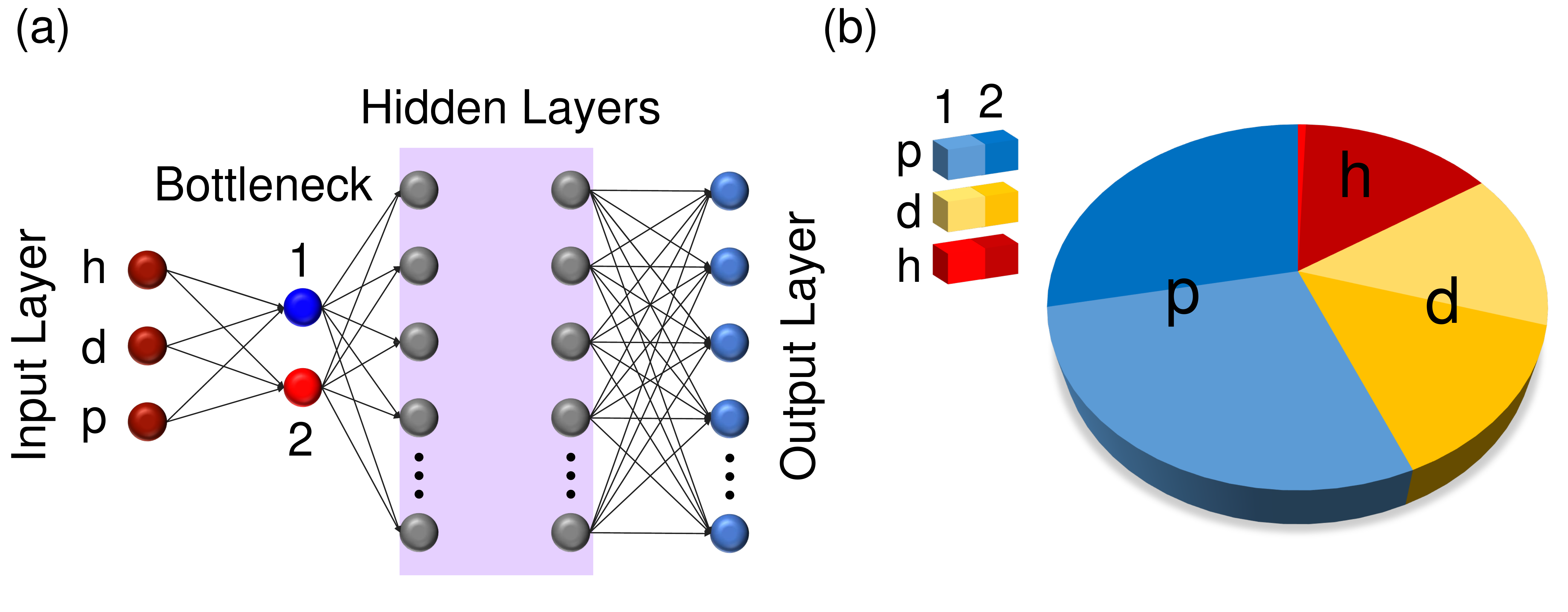}
\caption{(a) The architecture of the monolayer pseudo-encoder used to prioritize the most influential design parameter connections. (b) Strength of different design parameters connections to the bottleneck layer (the second layer with the minimum number of 2 neurons). $p$ has the dominant role in the design procedure.}
\label{fig:PseudoEnc}
\end{figure}

Figure~{\ref{fig:Structure}}(a) displays the all-dielectric MS made of HfO$_2$ NPs with height $h$ and diameter $d$ inside a square-lattice with periodicity $p$. The structure is normally illuminated by a TM-polarized plane wave of white light, and the reflectance response (associated with color) is measured at the far-filed. 

To effectively design the MS in Fig.~{\ref{fig:Structure}}(a) while avoiding the traditional computationally demanding optimization approaches (such as brute-force techniques or evolutionary approaches like genetic algorithms), we use a new deep-learning (DL)-based approach to predict the most influential design parameters in generating the desired sharp Fano-type resonant responses {\cite{kiarashinejad2019deep,YSMOA}}. The high level schematic of the deep-learning method is shown in Fig.~{\ref{fig:Blockdiagram}}. As the first step, we need a set of training data, which is reflectance spectra of 2000 different MSs with randomly selected design parameters, generated through full-wave simulations using finite-difference time-domain (FDTD) technique (Lumerical Inc.). Due to the periodic nature of the structure, the simulation domain is limited to one period ($p$) in the lateral directions (i.e., x and y in Fig.~{\ref{fig:Structure}}(a)) and perfectly matched layers are used on the top and bottom layers (in the z-direction in Fig.~{\ref{fig:Structure}}(a)). In all simulations, SiO$_\textrm{2}$ is assumed dispersionless with refractive index of 1.46. The optical constant of HfO$_\textrm{2}$ was measured in the visible spectrum by a Woollam M2000 Ellipsometer, as shown in Fig.~{\ref{fig:Structure}}(b), and added to the software library. To alleviate the computational cost, we reduce the dimensionality of the response space using an auto-encoder architecture. The auto-encoder is a neural-network (NN) that has the same input and output (i.e., in this case response space). The first part of the auto-encoder is an encoder which maps the original response space into the reduced response space (i.e., $\phi: R \longrightarrow F$). On the other hand, the decoder (i.e., $\psi :F \longrightarrow R$) is responsible to map the data from the reduced response space to the original response space.

To find the optimum dimensionality for the latent (or the reduced response) space we train a 11-layer neural network (200, 20, 20, 10, 10, $b$ , 10, 10, 20, 20, 200 nodes at each layer, respectively) for different sizes of the bottleneck layer (i.e., $b$). Note that the size of the original response space is 200. The activation function for all the hidden layers is tangent hyperbolic function, and there is no linear activation function for the input and output layers. Training the auto-encoder is performed using the backpropagation technique with minimizing the mean squared error (MSE); the optimizer is set to Adam. Figure~{\ref{fig:MSE}} represents the mean squared error (MSE) for different bottleneck sizes of the auto-encoder. Based on Fig.~{\ref{fig:MSE}}, we select the dimensions of the reduced response space to be 7. After training the auto-encoder, we reduce the dimensionality of the design space using a pseudo-encoder, which is a neural network that maps the design space to the response space through a bottleneck layer that corresponds to the reduced design space. In this work, we use a simple architecture in which the first hidden layer of the pseudo-encoder is the bottleneck layer with 2 nodes. The pseudo-encoder has 9 layers with 3, 2, 10, 10, 20, 20, 30, 30, 7 nodes in each layer, respectively, and the activation functions are all set to tangent hyperbolic except for the input and output layers. Training is performed by using the back-propagation method and Adam optimizer with minimizing the MSE. Since the trained toolkit compresses the underlying information of the design parameters in the bottleneck layer, the weights of the pseudo-encoder, especially in the first hidden layer, reveal the importance of each design parameter. Each input (i.e., design parameter) is normalized by subtracting from it the corresponding mean value and dividing the result by the standard deviation both calculated over the entire training dataset. No normalization I performed on the responses.

The trained pseudo-encoder in the DL-based approach provides valuable information in prioritizing and understanding the role of different design parameters. Figure~{\ref{fig:PseudoEnc}}(a) shows the schematic of the trained single-layer pseudo-encoder for the MS in Fig.~{\ref{fig:Structure}}(a) with three design parameters and a bottleneck layer (i.e., the second layer) with size 2 relating the design parameters to the response features of the structure. The weighting coefficients that relate the first (leftmost) and second layers of the pseudo-encoder in  Fig.~{\ref{fig:PseudoEnc}}(a) provide valuable information about the roles of different design parameters in the response of the structure. Each sector in Fig.~{\ref{fig:PseudoEnc}}(b) depicts the weights of the first layer of the pseudo-encoder to each node in the second (i.e., bottleneck) layer of Fig.~{\ref{fig:PseudoEnc}}(a) representing the role of each design parameter (i.e., $h$, $d$, and $p$) in forming the latent features. By comparing these weights in Fig.~{\ref{fig:PseudoEnc}}(b), it is clear that the periodicity (i.e., $p$) of the MS is the most dominant design parameter in changing the latent features and consequently manipulating the output response. In contrast, the NP height ($h$) has the least influence on the response. This is a very helpful observation as it allows the height to be selected by other constraints (e.g., the deposition thickness and the aspect-ration limitations in fabrication). Such valuable information cannot be obtained in using more conventional design and optimization techniques.

\begin{figure}[t]
\centering
\includegraphics[width=1\linewidth, trim={1cm 1cm 1cm 1cm},clip]{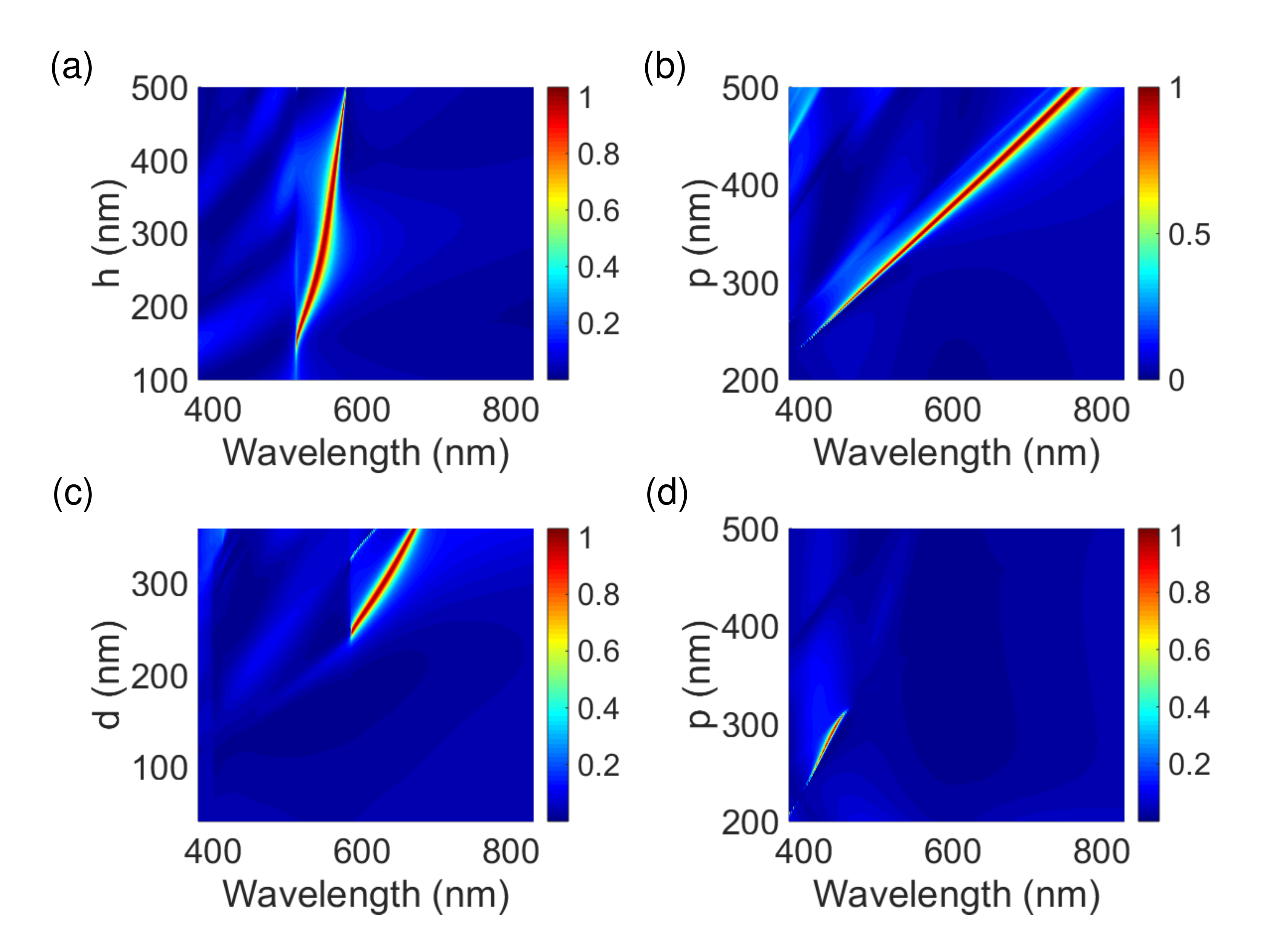}
\caption{Simulated reflectance profile of a unit cell of the MS shown in Fig.~{\ref{fig:Structure}}(a) versus (a) height of the constituents NPs, i.e., $h$, while other parameters are fixed at $p$ = 350 nm and $d = 0.75\,p$, (b) period of the unit cell, i.e., $p$, while the diameter is changing as $d = 0.75\,p$ and the height is fixed at $h$ = 350 nm, (c) diameter of the NPs, i.e., $d$, while $p$ = 400 nm and $h$ = 350 nm, (d) period of the unit cell while $d$ = 180 nm and $h$ = 350 nm.}
\label{fig:Sweep_DP}
\end{figure}

To confirm the results shown in Fig.~{\ref{fig:PseudoEnc}}(b), here we change the design parameters of the MSs and perform a series of simulations to analyze the sensitivity of the reflectance responses to design parameters. Figure~{\ref{fig:Sweep_DP}} shows the simulated reflectance amplitude profile of a unit cell of the MS shown in Fig.~{\ref{fig:Structure}} versus different structural parameters of the MS. Figure~{\ref{fig:Sweep_DP}}(a) shows that by changing the height of the NPs of the MS, the linewidth and spectral position of the corresponding Fano-type resonance slightly change. On the other hand, Fig.~{\ref{fig:Sweep_DP}}(b) shows that changing the periodicity of the MS and the diameter of constituent NPs results in a significant variation in both spectral position and linewidth of Fano-type resonances. Therefore, Figs.~{\ref{fig:Sweep_DP}}(a) and (b) confirm that $h$ has less significant role than $p$ and $d$ in changing the reflectance response, and in turn, changing the generated color, as we expect from our DL-based approach (see Fig.~{\ref{fig:PseudoEnc}}(b). To compare the importance of $p$ and $d$ in affecting the reflectance response of the MS, we first change $d$ and set $p$ = 400 nm and $h$ = 350 nm, and then change $p$ and fix $d$ = 180 nm and $h$ = 350 nm, and plot the resultant reflectance response in Figs.~{\ref{fig:Sweep_DP}}(c) and (d), respectively. We observe that the Fano-type resonances appear only for a limited range of $d$ and $p$. However, this range is smaller for the case of $p$ than that of $d$ confirming that the appearance of Fano-type resonances (i.e., the possibility of having high-quality colors) is more sensitive to the variation of $p$ than that of $d$.

\begin{figure}[b!]
\centering
\includegraphics[width=0.5\textwidth, trim={0.7cm 1cm 0cm 0cm},clip]{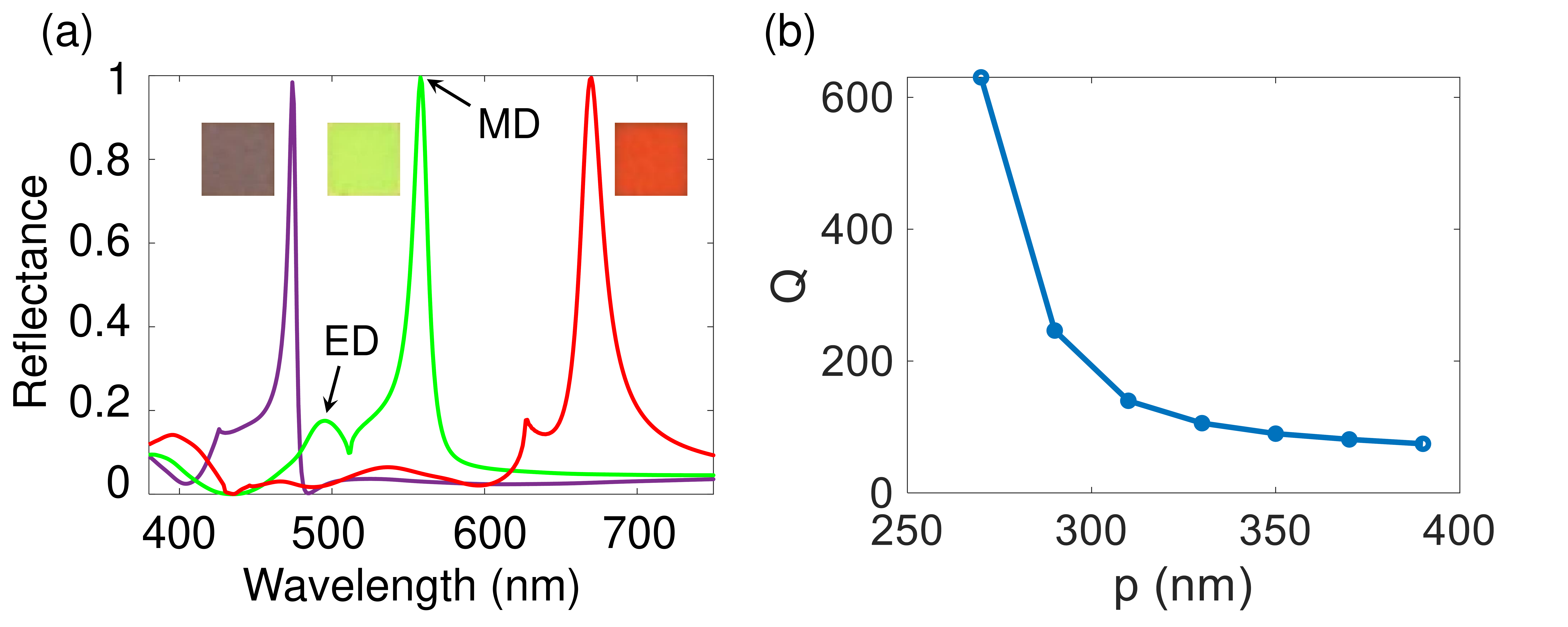}
\caption{The reflectance spectra for MSs in Fig.~\ref{fig:Structure}(a) designed for blue ($p$ = 270 nm), green ($p$ = 350 nm), and red ($p$ = 430 nm) colors. For all three structures, $d$ = 0.75p and $h$ = 350 nm. (b) $Q$ values for Fano-type resonances versus the periodicity of the MSs, i.e. $p$, where the diameters of the NPs are $d = 0.75\,p$.}
    \label{fig:Ref_Mie}
\end{figure}

As a result, by changing $p$, the spectral position of Fano-type resonances can be well tuned to induce different colors over the entire visible range. Figure~\ref{fig:Ref_Mie}(a) illustrates the reflectance spectra for three designed MSs each consisting of a square lattice of HfO$_2$ NPs with periodicities $p$ = 270 nm, 350 nm and 430 nm, diameters $d$ = 0.75 $p$, and all with a fixed height $h$ = 350 nm. The high purity of colors shown in Fig~{\ref{fig:Ref_Mie}}(a) is attributed to the high-$Q$ Fano-type resonances in the reflectance response. We estimate the $Q$ of Fano-type resonances by fitting the reflectance responses from MSs with standard Fano lineshape given by~{\cite{yang2014all}}:

\begin{equation}
R = |a_1 + i a_2 + \frac{b}{\omega - \omega_0 +i\gamma}|^2,
\label{eq:eq1}
\end{equation}

where $a_1$, $a_2$, and $b$ are the constant real numbers, $\omega_0$ is the central resonant frequency, and $\gamma$ is the overall damping rate of the resonance. The $Q$ is calculated by $Q=\omega_0/\gamma$. Figure~{\ref{fig:Ref_Mie}}(b) shows the evaluated $Q$ versus the diameter of the NPs for different MSs. We observe that by increasing the diameter of NPs, the $Q$ of Fano-type resonances decreases. However, the obtained $Q$s ($Q$ > 60) in this work are comparable with the best published results for Fano-type resonant color generation techniques using both plasmonic~{\cite{tseng2017two}} and all-dielectric~{\cite{shen2015structural}} MSs.


\begin{figure}[t]
\centering
\includegraphics[width=0.5\textwidth, trim={1cm 0cm 0cm 1cm},clip]{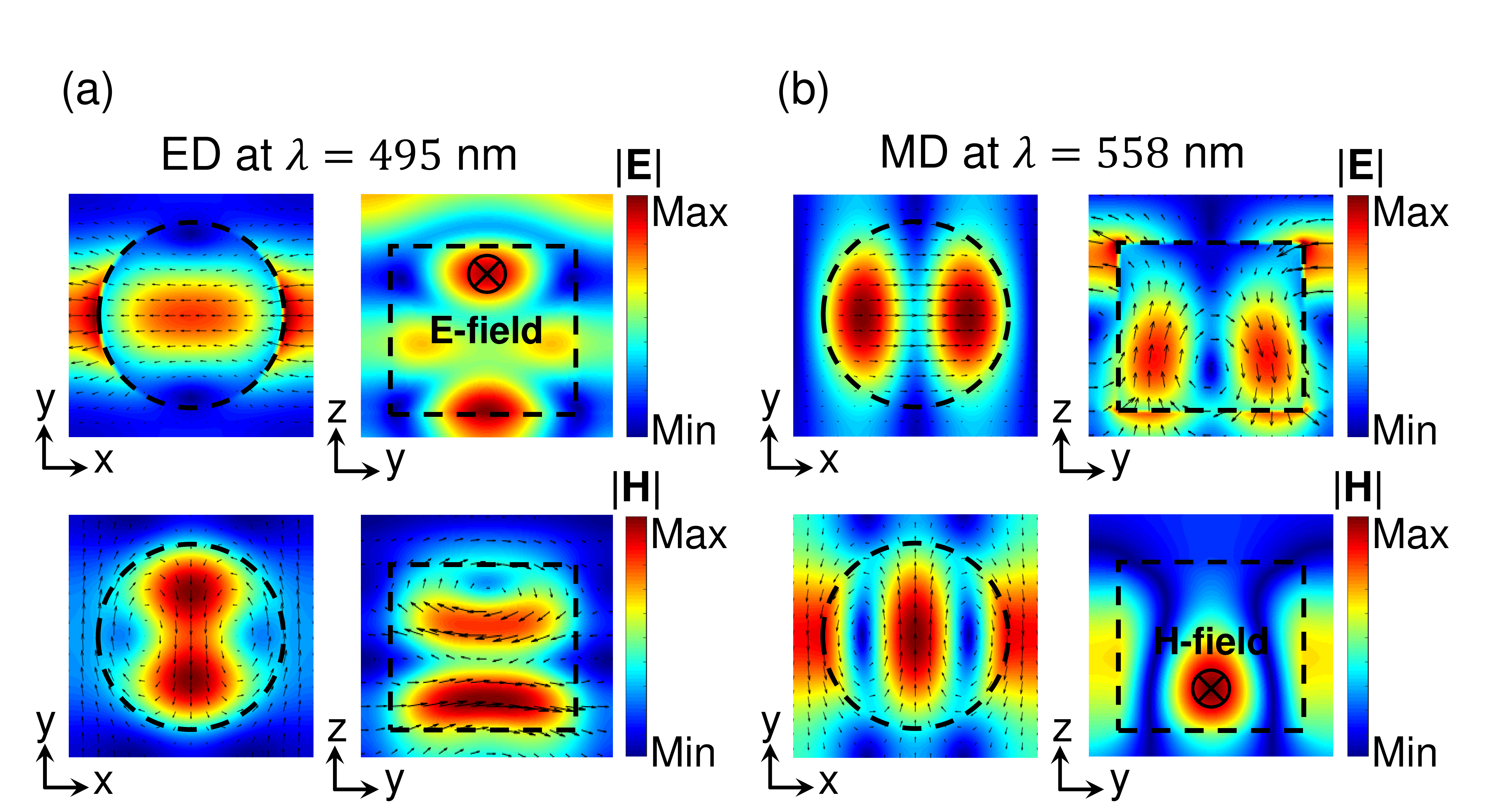}
\caption{Vector plots and color-coded distribution are shown for (a) electric and (b) magnetic fields inside the NP for the case of $p$ = 350 nm, $d$ = 0.75$p$, and $h$ = 350 nm. The structure supports the ED mode at $\lambda$ = 490 nm  (a) and the MD mode at $\lambda$ = 558 nm (b), which once coupled to the directly reflected light yields the sharp Fano-type peaks seen in Fig.~\ref{fig:Ref_Mie}(a).}
    \label{fig:MD_ED}
\end{figure}
\begin{figure}[b!]
\centering
\includegraphics[width=1\linewidth, trim={0cm 1cm 0cm 1cm},clip]{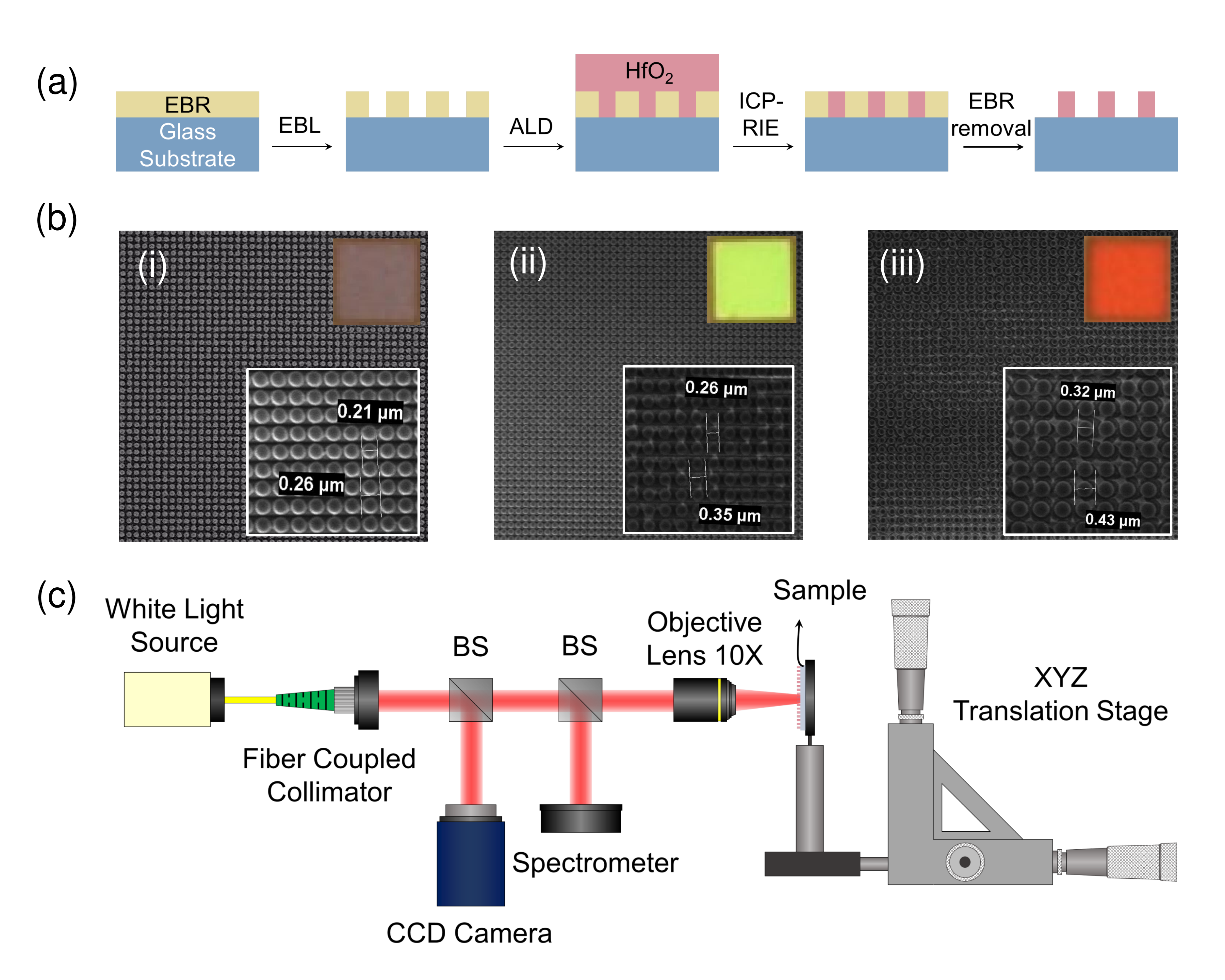}
\caption{(a) The fabrication process flow for implementing the colored all-dielectric MSs. (b) The SEM images of fabricated MSs designed for (i) blue ($p = 270$ nm), (ii) green ($p = 350$ nm), and (iii) red ($p = 430$ nm) colors, respectively. For all three structures, $d = 0.75 \, p$ and $h = 350$ nm. The measured diameter and periodicities are in good agreement with the designed ones. (c) Schematic of the characterization set-up for collecting the reflectance spectra of fabricated devices. The set-up consists of a white light source coupled to a fiber collimator to provide a collimated beam, two beam splitters (BSs), an objective lens, a spectrometer, and a CCD camera. The collimated beam is finally focused on the sample using a 10$\times$ magnification (NA = 0.1) objective lens, and the reflected light is then collected using the same lens.}
\label{fig:Fab_SEM_Setup}
\end{figure}

\begin{figure*}[t!]
\centering
\includegraphics[width=1\linewidth, trim={0cm 1cm 0cm 1cm},clip]{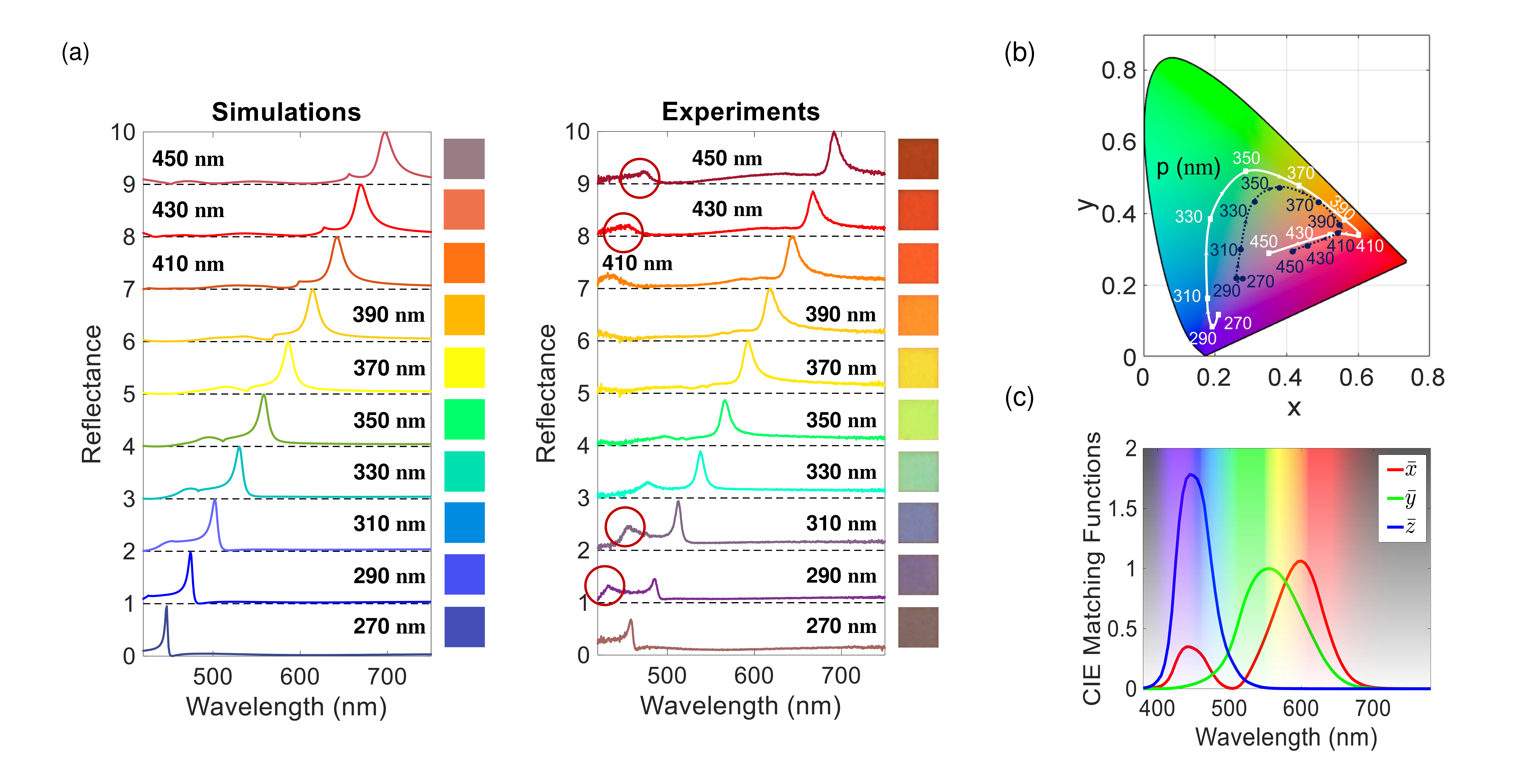}
\caption{(a) The simulation and experimental reflectance spectra and their corresponding colors for the structure in Fig.~\ref{fig:Structure}(a) with $h$ = 350 nm, $d$ = 0.75p and different periodicities (i.e., $p$). The red circles in the experimental results show spurious peaks due to fabrication imperfections. (b) CIE 1931 standard color-matching functions. (c) CIE color gamut formed by simulation (white squares) and experiment (blue dots) data in (a).}
\label{fig:RefColors_Gamut}
\end{figure*}

The sharp Fano-type resonances observed in Fig.~\ref{fig:Ref_Mie}(a) are primarily formed through the interaction of the confined narrow-band MD mode of the NPs with the wideband reflected mode from the structure. Figures~\ref{fig:MD_ED}(a) and \ref{fig:MD_ED}(b) show the electric ($|E|$) and magnetic ($|H|$) field patterns for the resonances at $\lambda$ = 495 nm and $\lambda$ = 558 nm, respectively, for the NPs with $p$ = 350 nm, $d=0.75 \,p$ and $h$ = 350 nm (corresponding to the green curve in Fig.~\ref{fig:Ref_Mie}(a)). Upon launching the white light into the MS, the constituent HfO$_2$ NPs, radiate at $\lambda$ = 495 nm like an ED recognized with the enhanced $|E|$ pattern and anti-parallel directions of $|H|$ at the opposite sides (i.e., top and bottom) of the NP seen in Fig.~\ref{fig:MD_ED}(a). However, the field pattern at $\lambda$ = 558 nm corresponds to a MD mode as seen in Fig.~\ref{fig:MD_ED}(b). This clearly shows the importance of the MD mode of the NP in the formation of the Fano-type resonances. To excite a MD in a NP, the direction of the incident electric field must experience a 180-degree change after transmission through the NP. It is well-known that this direction change happens when the effective wavelength of light inside the NP is comparable to the NP size, which is the case for our designed MSs. This is clearly seen from the $|E|$ pattern of the NP mode at $\lambda$ = 558 nm in Fig.~\ref{fig:MD_ED}(b). In the creation of this MD, a circular displacement current is formed inside the NP as seen from the electric-field lines in the $|E|$ pattern in Fig.~\ref{fig:MD_ED}(b). This circular displacement current can interact strongly with the incident light to induce a strong localized magnetic field at the center of the NP (as seen from the $|H|$ pattern in Fig.~\ref{fig:MD_ED}(b)). The coupling of the narrow-band MD mode of the NP with the broadband reflected white light results in the formation of the sharp Fano-type resonance that appears in the reflectance spectrum as observed in Fig.~\ref{fig:Ref_Mie}(a).

% \begin{figure}[t]
% \centering
% \includegraphics[width=0.5\textwidth, trim={1cm 0cm 0cm 1cm},clip]{Fig_Field_Profile.pdf}
% \caption{Vector plots and color-coded distribution are shown for (a) electric and (b) magnetic fields inside the NP for the case of $p$ = 350 nm, $d$ = 0.75$p$, and $h$ = 350 nm. The structure supports the ED mode at $\lambda$ = 490 nm  (a) and the MD mode at $\lambda$ = 558 nm (b), which once coupled to the directly reflected light yields the sharp Fano-type peaks seen in Fig.~\ref{fig:Ref_Mie}(a).}
%     \label{fig:MD_ED}
% \end{figure}
% \begin{figure}[b]
% \centering
% \includegraphics[width=1\linewidth, trim={0cm 1cm 0cm 1cm},clip]{Fab_SEM_Setup_Final.pdf}
% \caption{(a) The fabrication process flow for implementing the colored all-dielectric MSs. (b) The SEM images of fabricated MSs designed for (i) blue ($p = 270$ nm), (ii) green ($p = 350$ nm), and (iii) red ($p = 430$ nm) colors, respectively. For all three structures, $d = 0.75 \, p$ and $h = 350$ nm. The measured diameter and periodicities are in good agreement with the designed ones. (c) Schematic of the characterization set-up for collecting the reflectance spectra of fabricated devices. The set-up consists of a white light source coupled to a fiber collimator to provide a collimated beam, two beam splitters (BSs), an objective lens, a spectrometer, and a CCD camera. The collimated beam is finally focused on the sample using a 10$\times$ magnification (NA = 0.1) objective lens, and the reflected light is then collected using the same lens.}
% \label{fig:Fab_SEM_Setup}
% \end{figure}

Motivated by the physical intuition granted by the DL technique, several MSs were optimized to create different structural colors. Different aspects of three examples designed for blue, green, and red colors are shown in Fig.~\ref{fig:Ref_Mie}(a). To implement the designed MSs, we followed the step-by-step fabrication procedure shown in Fig.~\ref{fig:Fab_SEM_Setup}(a) \cite{RKN_PNAS}. First, a positive electron-beam resist (EBR) was spin-coated on a fused silica substrate and soft baked. Next, the electron-beam lithography (EBL) was performed to pattern the spin-coated EBR, which was followed by the developing process under gentle agitation. In the next step, a HfO$_2$ film was deposited inside the atomic layer deposition (ALD) chamber at 90$^{\circ} C$ (that was low enough to prevent the deformation of EBR patterns) using a standard two-pulse system of TEMAH and H$_2$O precursors. It is worth mentioning that ALD enabled us to create nanoscale features with minimum surface roughness as required in our designed MSs. Then, an etching process was carried out in an inductively-coupled-plasma (ICP) reactive-ion etching (RIE) system using a mixture of Cl$_{2}$ and BCl$_3$ gasses. Finally, the remaining EBR was exposed to the oxygen plasma and removed by overnight immersion of the sample inside the EBR solvent. Scanning electron microscopy (SEM) was performed to investigate the quality of the fabricated MSs. Figure~\ref{fig:Fab_SEM_Setup}(b) displays the SEM images of MSs with different designs and their corresponding colors (captured by an Olympus MX61 microscope) as shown in the insets.

To characterize the fabricated devices, we use a homemade optical set-up shown in Fig.~{\ref{fig:Fab_SEM_Setup}}(c). First, a fiber collimator is used to collimate the beam illuminated from a white light source. An objective (with 10$\times$ magnification, and numerical aperture (NA) of 0.1) is then used to focus the collimated incident beam onto the sample and collect the reflectance response as well. The reflected light passes through two beam splitters to be collected by a spectrometer and a charge-coupled device (CCD) camera for spectral measurement and imaging, respectively.

% \begin{figure*}[t]
% \centering
% \includegraphics[width=1\linewidth, trim={0cm 1cm 0cm 1cm},clip]{Sim_Exp_Gamut_Final.pdf}
% \caption{(a) The simulation and experimental reflectance spectra and their corresponding colors for the structure in Fig.~\ref{fig:Structure}(a) with $h$ = 350 nm, $d$ = 0.75p and different periodicities (i.e., $p$). The red circles in the experimental results show spurious peaks due to fabrication imperfections. (b) CIE 1931 standard color-matching functions. (c) CIE color gamut formed by simulation (white squares) and experiment (blue dots) data in (a).}
% \label{fig:RefColors_Gamut}
% \end{figure*}

\begin{figure}[!t]
\centering
\includegraphics[width=1\linewidth, trim={0.5cm 0cm 0.5cm 0cm},clip]{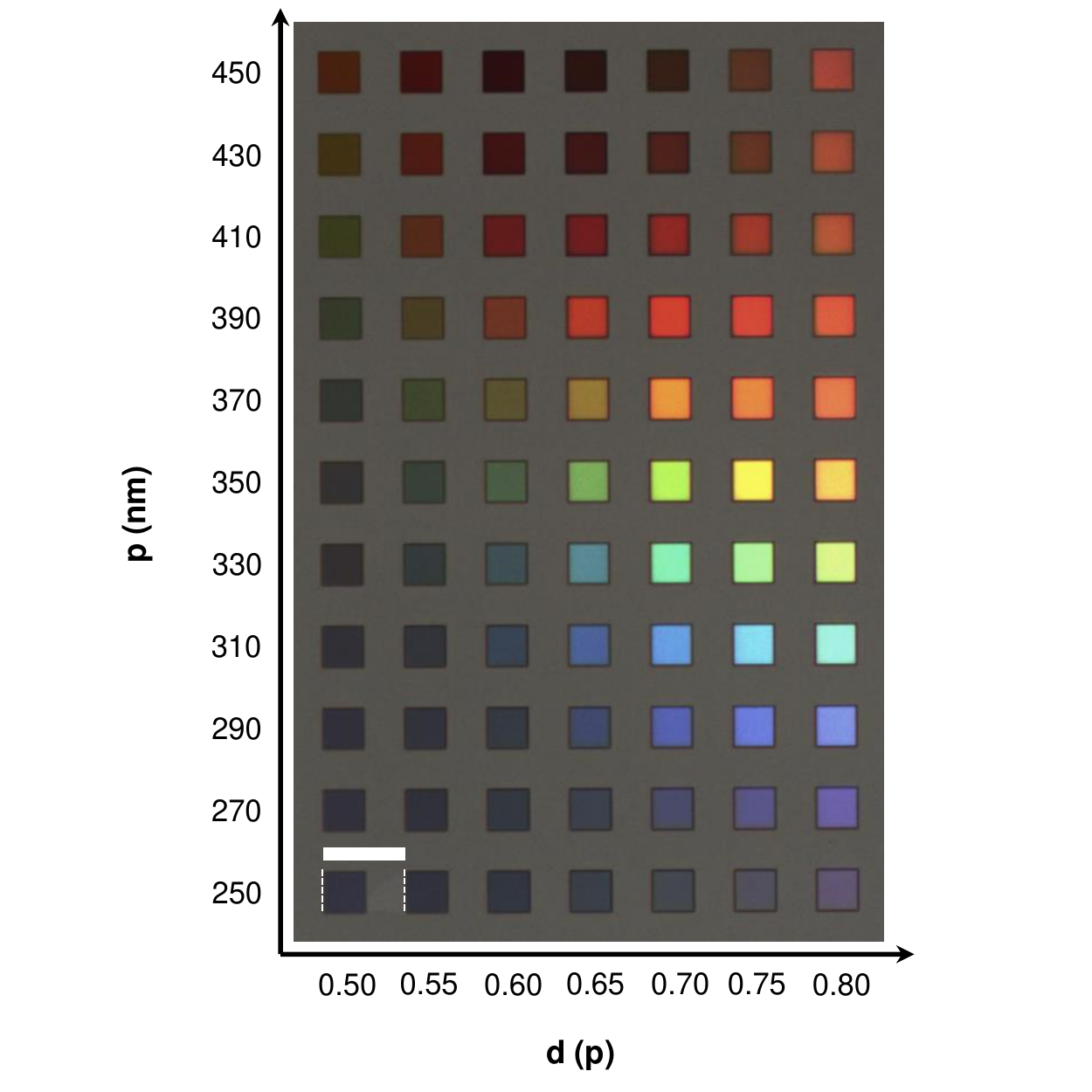}
\caption{The color palette for the fabricated array of all-dielectric MSs each composed of a 20$\times$20 $\mu$m$^2$ array of HfO$_2$ NPs. The image is captured by Olympus MX61 microscope through a 20$\times$ objective. The lattice periodicity and the diameter of NPs in each MS are changing from $p$ = 250 nm to $p$ = 450 nm (y-axis), and $d=0.50\,p$ to $d=0.80\,p$ (x-axis), respectively, while the height of NPs in all MSs are fixed to $h$ = 350 nm. The scale bar is 40 $\mu$m.}
\label{fig:Palette}
\end{figure}

\begin{figure}[!t]
\centering
\includegraphics[width=1\linewidth, trim={0cm 0cm 0cm 0cm},clip]{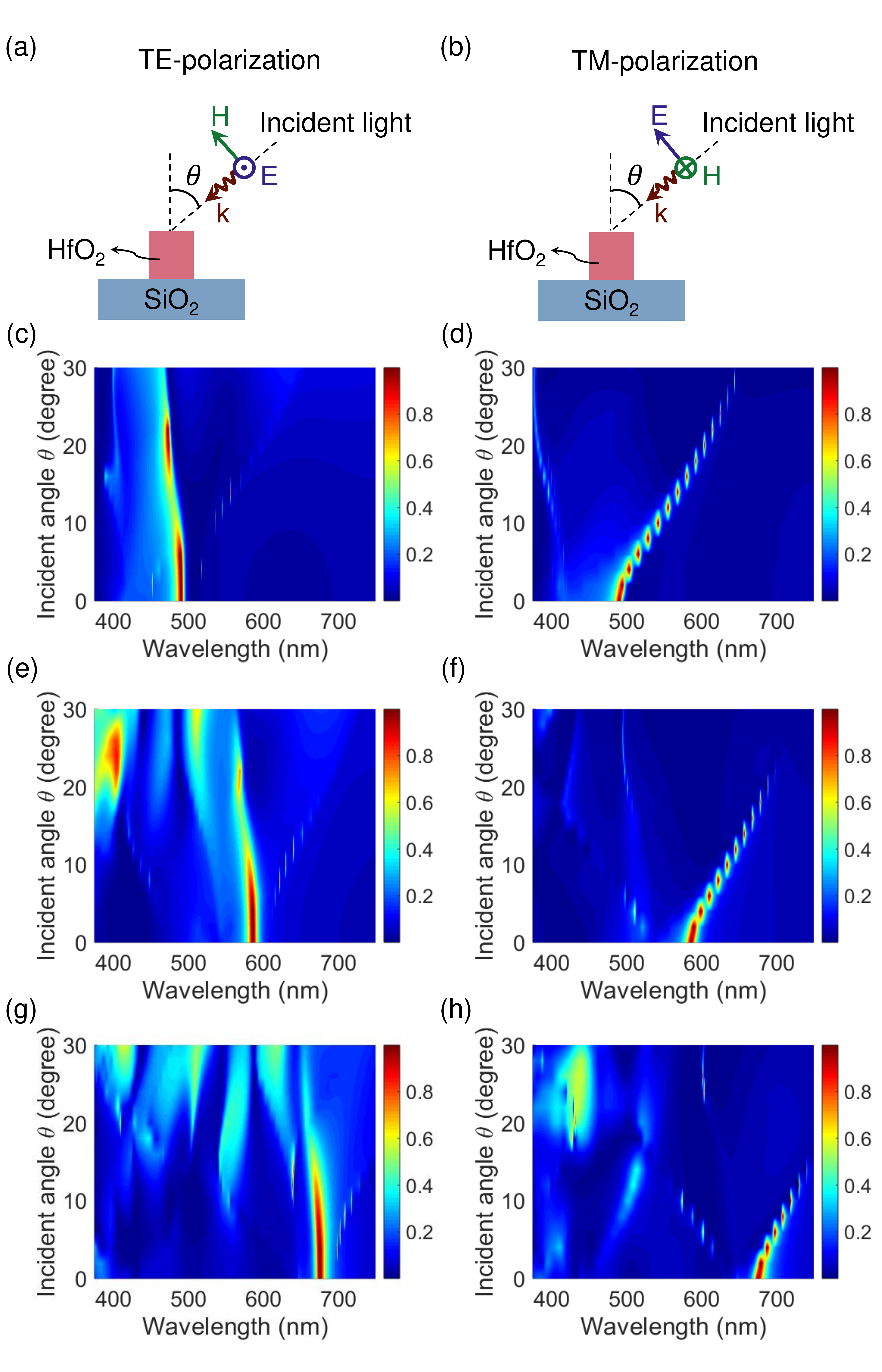}
\caption{The reflectance spectra sensitivity to the incident angle $\theta$ for (a, c, e, g) TE-polarized (electric field parallel to the interface) and (b, d, f, h) TM-polarized (magnetic field parallel to the interface) incident light. (c, d), (e, f) and (g, h) show this incident angle sensitivity of the reflected light from the MS designed for blue color (i.e. $p=$290 nm), green color (i.e. $p=$350 nm) and red color (i.e. $p=$430 nm), respectively. The height of HfO$_2$ NPs are fixed at $h=$ 350 nm for all cases.}
\label{fig:TE_TM_Angle}
\end{figure}

The simulated and measured reflectance spectra for the designed MSs along with their resulting colors are shown in Fig.~\ref{fig:RefColors_Gamut}(a), where the periodicity of the lattices are changing from $p = 270$ nm to $p$ = 450 nm, while the diameter and the height of NPs are chosen as $d$ = 0.75p, and $h$ = 350 nm, respectively. A good agreement is observed between simulation results and those obtained via experiment, which confirms the reliability of the fabrication process. To study the generated colors, the International Commission on Illumination (CIE) XYZ tristimulus values corresponding to the reflectance spectra are calculated as
\begin{gather}
X = \frac{1}{k} \int{I(\lambda) R(\lambda) \bar{x}(\lambda) d\lambda},\nonumber \\
Y = \frac{1}{k} \int{I(\lambda) R(\lambda) \bar{y}(\lambda) d\lambda},\\
Z = \frac{1}{k} \int{I(\lambda) R(\lambda) \bar{z}(\lambda) d\lambda}.\nonumber 
\label{eq:eq2}
\end{gather}
Here, $k$ is the normalization factor, \(I(\lambda)\) is the spectral energy distribution of the reference light, \(R(\lambda)\) is the far field reflectance spectrum obtained from the designed MS under illumination, \(\bar{x}(\lambda)\), \(\bar{y}(\lambda)\), and \(\bar{z}(\lambda\)) are the CIE 1931 standard color-matching functions (shown in Fig.~\ref{fig:RefColors_Gamut}(b)) \cite{a_Si_disk}. These chromaticity functions are then normalized as $x=X/(X+Y+Z)$ and $y=Y/(X+Y+Z)$, which fall between 0 and 1, to represent the colors in the CIE 1931 chromaticity diagram (also known as the color gamut) as shown in Fig.~\ref{fig:RefColors_Gamut}(c).

\begin{figure*}[t!]
\centering
\includegraphics[width=1\linewidth, trim={1cm 1cm 1cm 1cm},clip]{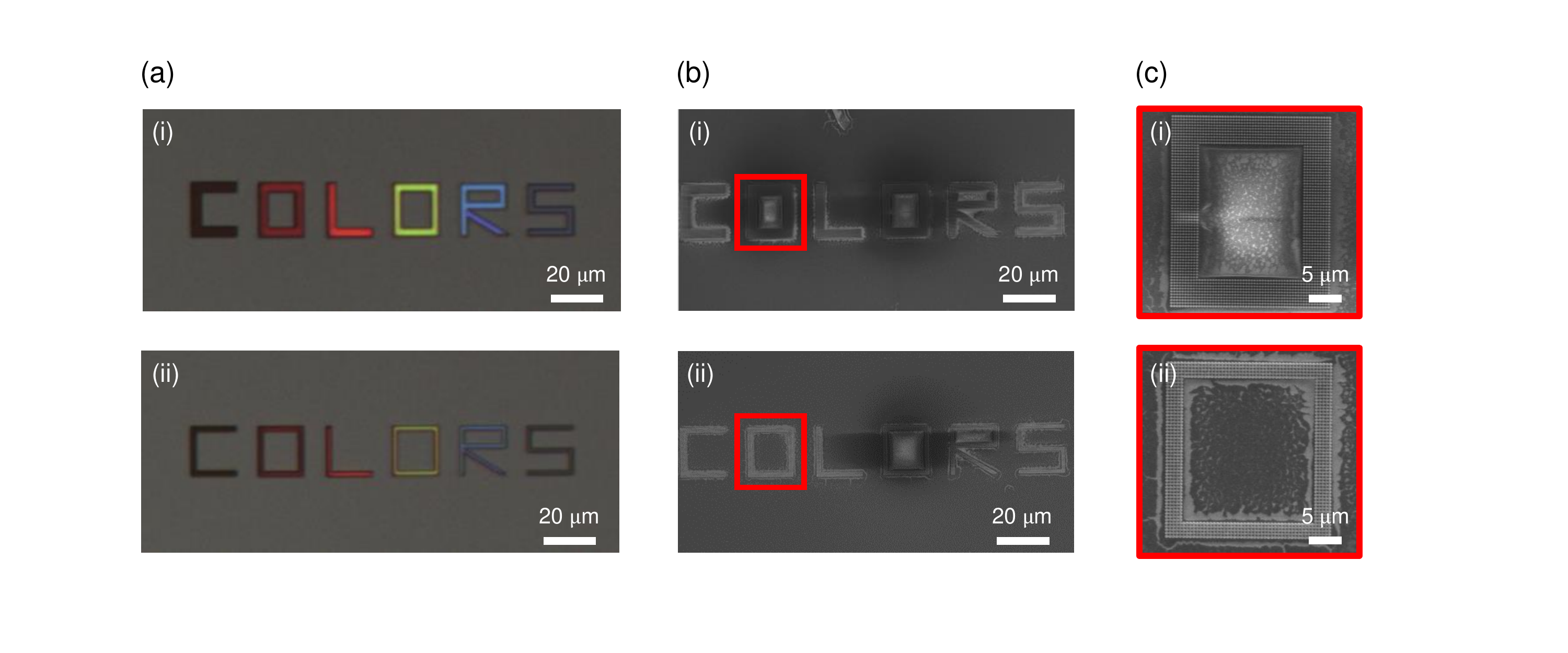}
\caption{Effect of color pixel size on generated colors. (a) The images of the characters "COLORS" with color pixels consist of (i) 10 and (ii) 5 HfO$_2$ NPs. The images are captured by Olympus MX61 microscope through 20$\times$ objective. (b, c) The SEM images of fabricated characters. The shade effects in (b) and (c) are due to the residue of spin-coated Espacer as a conductive layer facilitating the SEM imaging process.}
\label{fig:COLORS_SEM}
\end{figure*}

The results in Fig.~\ref{fig:RefColors_Gamut}(c) confirm that a wide range of structural colors can be achieved by our MSs.  Figure~\ref{fig:RefColors_Gamut}(c) shows a good agreement between simulation and experimental data for yellow, orange and red colors (achieved with 330 nm < $p$ < 410 nm) while some differences between the two types of results are observed for green, blue, and dark red colors ($p$ < 330 nm and $p$ > 410 nm).  We believe these differences are primarily caused by fabrication imperfections. The high current intensity used in the EBL process to form the structures with higher aspect ratios (i.e., smaller $d$ and $p$ < 330 nm) causes an error in the diameter of the NPs (and thus, in the observed color). On the other hand, the long ALD process for the formation of the HfO$_2$ layer results in the hard-baking of the EBR, which is difficult to remove. Thus, a residue of the EBR appears on top of the NPs (see Fig.~\ref{fig:Fab_SEM_Setup}(b)) resulting in the modification of the reflectance spectra. This is clearly observed in Fig.~\ref{fig:RefColors_Gamut}(a) in the form of smaller spurious peaks on the left side of the main peak of each color in the experimental results (shown by red circles), especially in the structures with $p$ < 330 nm and $p$ > 410 nm. Once multiplied by  \(\bar{x}(\lambda)\), \(\bar{y}(\lambda)\), and \(\bar{z}(\lambda\)) (see Fig. \ref{fig:RefColors_Gamut}(b)), these spurious peaks result in changing the values of $x$  and $y$ (both obtained from Eq.(2)) and a shift in the color. The amount of this shift (or discrepancy between simulation and experimental results) primarily depends on the locations of the main and the spurious reflectance peaks. It is highest for blue colors where the main peak and the spurious peak overlap (i.e., the blue color changes to a more brown color at $p$ = 270 nm as seen from Fig.~\ref{fig:RefColors_Gamut}(a)). As a result, the overall color gamut is Fig.~\ref{fig:RefColors_Gamut}(c) covers a smaller area compared to the theoretical predictions. The development of new processes for avoiding the EBR in the fabrication of the NPs to solve this discrepancy is currently underway. Finally, the non-ideal operation of lower-wavelength filters installed in the camera  reduces the color purity in the blue region. 

Figure~{\ref{fig:Palette}} shows the bright field image of a full color palette generated by an 11 $\times$ 7 array of fabricated MSs under unpolarized white light illumination through a 20$\times$ objective (NA = 0.45). The height of NPs in all MSs are fixed to $h$ = 350 nm, while the periodicity in each MS is varying from $p$ = 250 nm to $p$ = 450 nm in 20 nm-step (y-axis in the palette in Fig.~{\ref{fig:Palette}}), and the diameter of NPs in each MS is changing from $d$ =0.5p to $d$ =0.8p (x-axis in the palette in Fig.~{\ref{fig:Palette}}). Figure~{\ref{fig:Palette}} shows that a wide range of vivid and high quality colors can be obtained using the MSs composed of HfO$_2$ NPs. It also shows that the colors change from deep blue to dark red by increasing $p$ when $d/p$ is larger than 0.6.

To study the sensitivity of the reflectance spectra to the incident angle, we performed full-wave simulations by changing the incident angle $\theta$ (Fig.~{\ref{fig:TE_TM_Angle}}(a, b)) from 0 to 30 degrees, and plot a 2D reflectance map as a function of both wavelength and $\theta$ in Fig.~{\ref{fig:TE_TM_Angle}}. The geometrical parameters of MSs are $p$ = 290 nm (blue color), $p$ = 350 nm (green color) and $p$ = 430 nm (red color), in Fig.~{\ref{fig:TE_TM_Angle}}(c, d), (e, f), and (g, h), respectively. As Figs.~{\ref{fig:TE_TM_Angle}}(c, e, g) show, for the case of TE-polarization (electric field parallel to the interface) the spectral positions of reflectance peaks are almost insensitive to the incident angle. The reason is that, these peaks primarily originate from the MD mode inside the HfO$_2$ NP, and this mode itself, as previously discussed, is ascribed to the electric field component parallel to the top surface of the HfO$_2$ NPs. For the case of TM-polarization (magnetic field parallel to the interface), we observe that increase in incident angle from 0 to 30 degrees not only red-shifts the spectral positions of the reflectance peaks (Fano-type resonances), but also damps the reflectance amplitude (Figs.~{\ref{fig:TE_TM_Angle}}(d, f, h)). The reason for this angle-sensitivity for the case of TM-polarization is that by increasing the incident angle, the electric field component parallel to the top of the HfO$_2$ NP is mitigated, which consequently results in weakening of the MD mode. Therefore, the efficiency of the Fano-type resonances (which are attributed to the constructive interference of MD mode with the reflected light) decreases slightly.

\begin{figure*}[t!]
\centering
\includegraphics[width=1\linewidth, trim={0.5cm 1.5cm 0cm 1cm},clip]{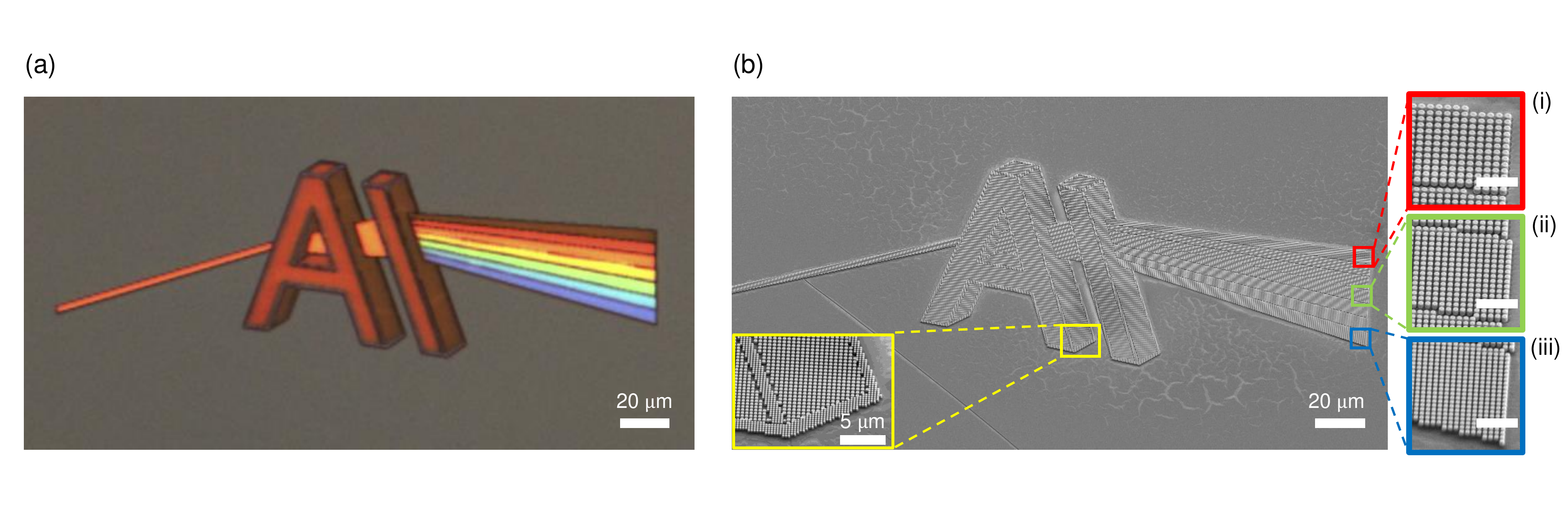}
\caption{Demonstration of a full-color image reproduction. (a) The reproduced image of a AI-prism logo captured by Olympus MX61 microscope through 20$\times$ objective. The colors in the reproduction are generated by HfO$_2$ NPs with varying lattice periodicities (i.e. from $p=$270 nm for violet color,  to $p=$430 nm for red color), and diameter (i.e $d=0.75\,p$) as shown in the SEM images in corresponding insets (i), (ii) and (iii). The scale bar in (i), (ii) and (iii) images is 2 $\mu$m. The inset in the left bottom of (b) shows that using color pixels consisting of 5 NPs, the outline color (violet here) can be generated.}
\label{fig:AI_prism}
\end{figure*}

The color pixel size in Fig.~{\ref{fig:Palette}} is 20$\times$20 $\mu$m$^2$. To study the effect of the color pixel size on the quality of the generated colors on one side and determine the resolution limit in our color printing technique on the other side, we generate an image of the characters "COLORS" (as shown in Fig.~{\ref{fig:COLORS_SEM}}) with color pixels of different sizes. In Fig.~{\ref{fig:COLORS_SEM}}, each color pixel contains 10 and 5 NPs for (i) and (ii), respectively. It is also worth mentioning that all the images in Fig.~{\ref{fig:COLORS_SEM}} are captured by an Olympus MX61 microscope through a 20$\times$ objective with NA = 0.45. For images captured with objectives with larger NA, the grating modes can be excited due to the oblique incidence with large incident angles which degrade the brightness of the generated colors.

To show the practical applicability of our color printing technique, we designed and fabricated an AI-prism logo (see Fig.~{\ref{fig:AI_prism}}(a)) containing a wide range of colors. To confirm that the designed logo was successfully fabricated, the SEM images are shown in Fig.~{\ref{fig:AI_prism}}(b).

To obtain even a larger color gamut with higher saturation, one can consider HfO$_2$ MSs with more complex unit cells and lattice structures or even MSs in hybrid multi-layer material platform. Our DL-based design technique can be used for designing such complex structures for which the conventional design techniques cannot be used for excessive computation requirements. Also, the valuable intuitive understanding of the roles of design parameters obtained by our DL-based technique can considerably facilitate the design process for complex structures.

\section*{Conclusions}
In summary, we demonstrated here an all-dielectric MS consisting of a square-lattice of NPs of HfO$_2$ to generate different structural colors. Such a capability is attributed to the strong induced ED and MD resonance modes of the engineered NPs. The coupling of MD mode to the directly reflected white light leads to narrow-band Fano-type resonances, which result in a wide color gamut. Furthermore, due to the near-zero loss of HfO$_2$, the quality of the generated colors are highly vivid and pure. In this creation, our DL-based analysis and optimization technique was helpful in reducing the computation complexity while providing intuitive information about the roles of different design parameters. 

\section*{Conflicts of interest}
There are no conflicts to declare.

%%%END OF MAIN TEXT%%%


\balance


%%%REFERENCES%%%
\bibliography{References} %You need to replace "rsc" on this line with the name of your .bib file
\bibliographystyle{rsc} %the RSC's .bst file

\end{document}